\documentclass[twocolumn,superscriptaddress]{revtex4-2}

\usepackage[dvipsnames]{xcolor}
\usepackage{hyperref}
\hypersetup{
  breaklinks=true,
  colorlinks=true,
  allcolors=BlueViolet,
}

\usepackage{graphicx}
\graphicspath{{./figures/}} % set path for all figures

\usepackage[yyyymmdd]{datetime} % set date format
\usepackage[inline]{enumitem} % for in-line lists
\setlist[enumerate]{label=(\arabic*)} % default numbering format

\usepackage{physics,braket,bm,amssymb} % physics and math
\usepackage{pifont} % for \cmark and \xmark
\newcommand{\yes}{\textcolor{Green}{\ding{51}}}
\newcommand{\no}{\textcolor{Red}{\ding{55}}}
\newcommand{\maybe}{\textcolor{Orange}{?}}

\newcommand{\Infleqtion}{Infleqtion, Inc., Chicago, IL 60604, USA}
\newcommand{\CSL}{CSL Behring, Bradley, IL, 60915, USA}

\usepackage{comment}

\begin{document}

\title{Sphere Packing on a Quantum Computer for Chromatography Modeling}
\author{Benjamin Hall}
\email{co-first author: ben.hall@infleqtion.com}
\affiliation{\Infleqtion}
\author{Ian Njoroge}
\email{co-first author: ian.njoroge@cslbehring.com}
\affiliation{\CSL}
\author{Colin Campbell}
\affiliation{\Infleqtion}
\author{Bharath Thotakura}
\affiliation{\Infleqtion}
\author{Rich Rines}
\affiliation{\Infleqtion}
\author{Victory Omole}
\affiliation{\Infleqtion}
\author{Maen Qadan}
\affiliation{\CSL}
\date{\today}

\begin{abstract}

Column chromatography is an important process in downstream biopharmaceutical manufacturing that enables high-selectivity separation of proteins through various modalities, such as affinity, ion exchange, hydrophobic interactions, or a combination of the aforementioned modes. Current mechanistic models of column chromatography typically abstract particle-level phenomena, in particular adsorption kinetics. 
A mechanistic model capable of incorporating particle-level phenomena would increase the value derived from mechanistic models. 
To this end, we model column chromatography via sphere packing, formulating three versions, each with increasing complexity.
The first, homogeneous circle packing, is recast as maximum independent set and solved by the Quantum Approximate Optimization Algorithm on a quantum computer. 
The second, heterogeneous circle packing, is formulated as a graphical optimization problem and solved via classical simulations, accompanied by a road map to a quantum solution. 
An extension to the third, heterogeneous sphere packing, is formulated mathematically in a manner suitable to a quantum solution. 
Finally, detailed resource scaling is conducted to estimate the quantum resources required to simulate the most realistic model, providing a pathway to quantum advantage.
\end{abstract}

\maketitle

%%%%%%%%%%%%%%%%%%%%%%%%%%%%%%%%%%%%%%%%%%%%%%%%%%%%%%%%%%%%%%%%%%%%%%

\section{Introduction}

As quantum computational power continues to grow rapidly \cite{willow}, it is paramount that industries which rely heavily on computation start to plan today for tomorrow's quantum advantage. 
One such industry is the biopharmaceutical industry, which uses massive amounts of classical computation to aid everything from drug discovery to operational optimization \cite{comp_pharm}. 
In this work, we investigate the utility of quantum computing to aid in the mechanistic modeling of column chromatography. 
Column chromatography is a primary separation method in downstream biopharmaceutical manufacturing\cite{Rathore2018, Kumar2020}. 
The modeling of chromatography columns has been utilized in accelerating process development and process scale-up as well as to maximize process productivity\cite{Kumar2020, Shekhawat2019, Borg2014, Espinoza2024}. Mechanistic models, which incorporate mathematical models of the transport and adsorption processes present in column chromatography, provide greater accuracy and precision in comparison to empirical models based mainly on statistical methods\cite{Kumar2020, Shekhawat2019, Borg2014, Espinoza2024}. 
Current approaches to mechanistic modeling typically abstract the particle-level adsorption kinetics through the use of generic isotherms\cite{Kumar2020, Borg2014, Espinoza2024}. 

Here, we begin the process of modeling particle-level phenomena in chromatography columns.
It is optimal for the chromatography separation process to generate tightly packed chromatography columns without affecting the integrity of the individual particles.
This problem is modeled as bounded sphere-packing, in which spheres are packed to minimize empty space.
Bounded sphere packing is further cast as a graphical optimization problem amenable to quantum solutions.

\begin{figure}[b]
    \centering
    \includegraphics[width=1\linewidth]{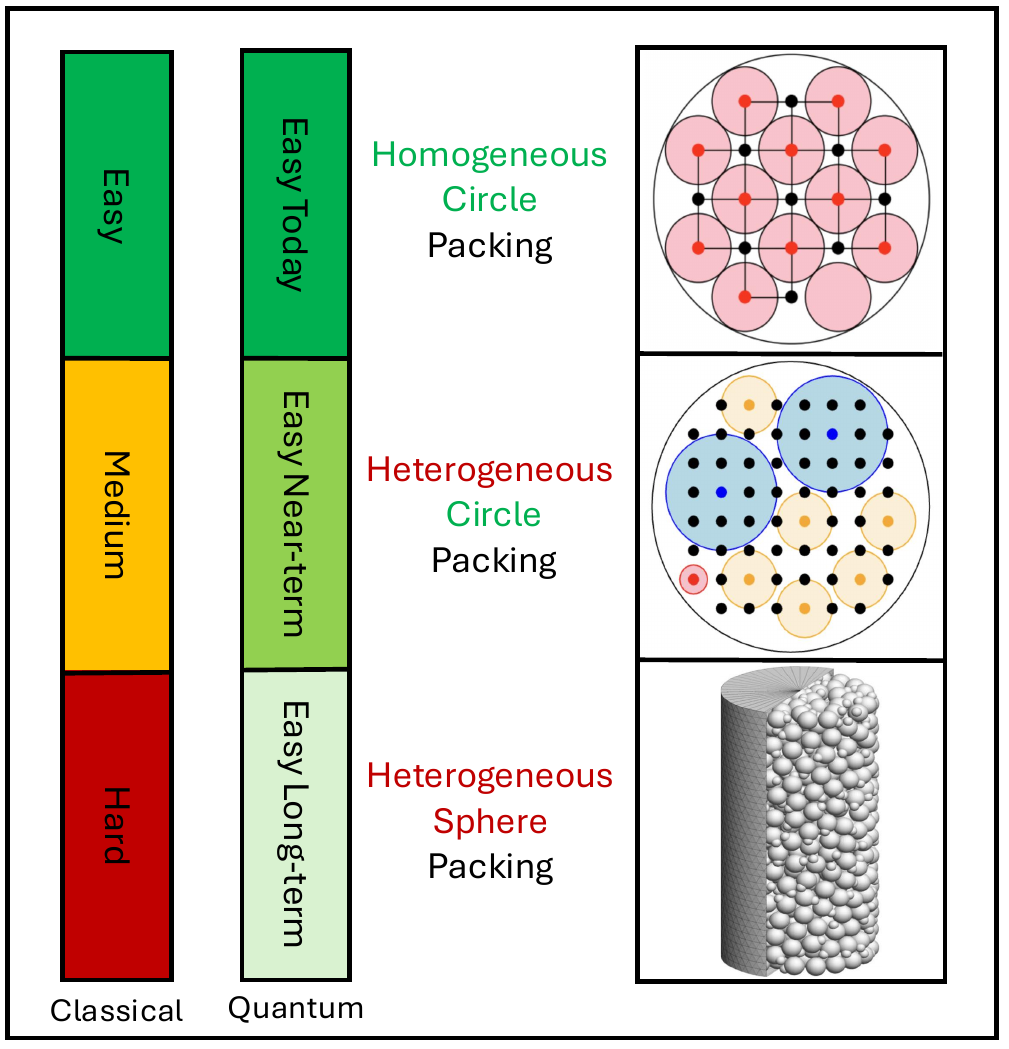}
    \caption{Three complexity levels of sphere packing for modeling chromatography: homogeneous circle packing, heterogeneous circle packing, and heterogeneous sphere packing, along with their respective difficulty levels for classical and quantum solutions over time.}
    \label{fig:figure1}
\end{figure}

We break the problem of sphere packing into three levels of increasing complexity: homogeneous circle packing, heterogeneous circle packing, and heterogeneous sphere packing (Figure \ref{fig:figure1}).
We solve example of homogeneous circle packing on real quantum hardware that exists today. 
Details of the hardware experiment include discussions on hyperparameter optimization, compilation, and noisy simulation.
Evidence of parameter concentration is also included. 
Numerical results for heterogeneous circle packing via classical computation were executed, and the problem was formulated in a manner suitable for quantum computation. 
Finally, heterogeneous sphere packing is shown to be an extension of the heterogeneous circle packing formulation, and resource estimation is performed to estimate the quantum resources required to solve it.
We will argue that while classical computers can approximately solve all three formulations today, their difficulty in doing so scales exponentially with increasing problem size and formulation complexity.
And in contrast, while quantum computers can only solve the simplest formulation today, there is good evidence that in time they will be able to solve all three levels of complexity, each with only polynomially scaling difficulty.

\begin{figure}[t]
    \centering
    \includegraphics[width=1\linewidth]{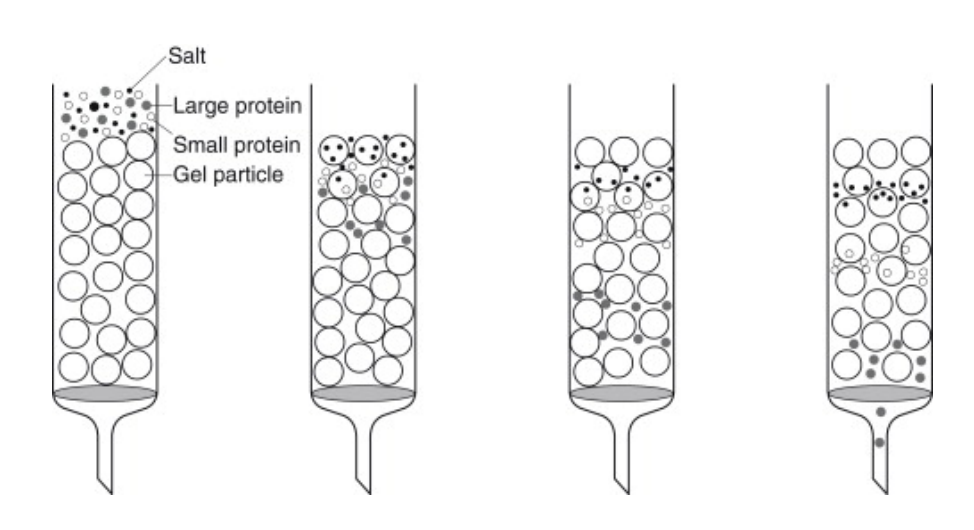}
    \caption{Schematic of chromatography for protein filtration.
Different proteins filter through the column of spherically modeled gels at different rates, leading to their separation.}
    \label{fig:chroma}
\end{figure}

\subsection{Sphere Packing}

\begin{table}[b]
  \centering
  \begin{tabular}{cl|c|c}
    \multicolumn{2}{c|}{unbounded} &
    \begin{tabular}{c} homogeneous \end{tabular} &
    \begin{tabular}{c} heterogeneous \end{tabular}   \\ \hline\hline
    2 dimensional & & \yes & \maybe \\ \hline
    3 dimensional & & \yes & \maybe \\ \hline
  \end{tabular}
  \caption{Unbounded homogeneous/heterogeneous sphere packing in two and three dimensions. \yes \ indicates the existence of a general analytical solution while \maybe \ indicates the existence of only partial analytical solutions.}
  \label{tab:unbounded}
\end{table}

\begin{table}[b]
  \centering
  \begin{tabular}{cl|c|c}
    \multicolumn{2}{c|}{bounded} &
    \begin{tabular}{c} homogeneous \end{tabular} &
    \begin{tabular}{c} heterogeneous \end{tabular}   \\ \hline\hline
    2 dimensional & & \maybe & \no \\ \hline
    3 dimensional & & \maybe & \no \\ \hline
  \end{tabular}
  \caption{Bounded homogeneous/heterogeneous sphere packing in two and three dimensions. \maybe \ indicates the existence of only partial analytical solutions. \no \ indicates no analytical solutions.}
  \label{tab:bounded}
\end{table}

As mentioned above, we will focus on the packing aspect of protein chromatography. 
The chromatography resin can be modeled as porous spheres which are to be packed into a cylindrical column as tightly as possible, meaning in a way that minimizes empty space (Figure \ref{fig:chroma}). This is the mathematical problem known as sphere packing. There are three features that can be toggled when discussing sphere packing: homogeneity, boundedness, and dimension: The spheres' radii can be homogeneous (all the same) or heterogeneous (different). The problem can be unbounded (the packer has unlimited space available) or bounded (the packer has limited space available). Finally, the problem can be solved in two dimensions (circle packing) or three dimensions (sphere packing). While we consider both homogeneous and heterogeneous cases, we only consider bounded cases as chromatography is bounded by the filtration cylinder. Finally, while sphere packing can, in principle \cite{hyper}, be formulated in any dimension, we start with two dimensions for simplification and later give a path to the practical case of three dimensions.

Unbounded homogeneous packing has a long history. In 1773, Joseph Lagrange proved that the optimal circle packing is the hexagonal packing arrangement \cite{lagrange_proof}, which achieves a packing density of $\pi/(2\sqrt{3})\approx0.91$. As for sphere packing, in 1611 Johannes Kepler conjectured that cubic close packing and the hexagonal close packing are optimal in their density of $\pi/(3\sqrt{2})\approx0.74$. His conjecture was finally proven, with the aid of computer proof assistants, in 2017 \cite{kepler_conjecture}. Much less is known about unbounded heterogeneous packing except for special ratios of radii for two sphere sizes \cite{nonuniform_bounded_circle}, leaving a general analytical solution an open problem within the field. The existence of general or analytical solutions is summarized for unbounded packing in Table \ref{tab:unbounded}.

However, as we are modeling protein chromatography, we must deal exclusively with bounded packing. In terms of homogeneous packing, the optimal circle packing within a circle is known for several small numbers of circles (and conjectured for others) when the radius of the circles is fixed and the radius of the boundary circle is allowed to be arbitrary.
It is an ongoing area of research as, for example, the optimal packing for 14 circles was only proven in 2024 \cite{14_circles}.
Homogeneous sphere packing within a cylinder has been studied via simulated annealing \cite{Mughal_2012, Chan_2011, Fu_2016} but general optimal solutions are not known. However, for homogeneous sphere packing within a cube, partial analytical solutions are known for certain small numbers of spheres \cite{spheres_in_cube}. Finally, the problem of bounded, heterogeneous sphere packing is the least well studied, for which general optimal solutions are also unknown. The existence of partial analytical solutions or no analytical solutions is summarized for bounded packing in Table \ref{tab:bounded}. A realistic model for protein chromatography is bounded, heterogeneous sphere packing, which has no known general or even partial analytical solutions.
This makes our problem a good candidate for quantum computation, whose utility will be argued for in the Resource Scaling section (\ref{resource_scaling}).

\subsection{QAOA}

To be solved computationally, the continuous problem of sphere packing must be discretized, leading to a discrete optimization problem. The problem can further be reduced to an integer optimization problem, as detailed in Subsection \ref{src_formulation}. Depending on the constraints, the search space of such problems can grow exponentially with problem size. Additionally, they are often NP-hard, meaning there is no known classical algorithm that can solve them in polynomial time. One promising near-term quantum algorithm that has been proposed to tackle discrete optimization problems is QAOA: the Quantum Approximate Optimization Algorithm \cite{qaoa}. QAOA is a variational hybrid quantum algorithm - requiring the tuning of variational parameters to minimize a cost function and the utilization of both a quantum and classical computer.

The algorithm is based on the quantum adiabatic theorem which states that if a system starts in the ground state of an initial Hamiltonian and evolves slowly enough to a final Hamiltonian, then the system will end in the ground state of said final Hamiltonian. This provides a useful computational tool for discrete optimization problems: simply start in the ground-state of an easy to solve Hamiltonian and slowly evolve to a final Hamiltonian which encodes the minimization problem. According to the quantum adiabatic theorem, the system will end in the ground state of this final Hamiltonian, which will encode the solution to the minimization problem (the state that minimizes the cost function). QAOA is a discretized version of the quantum adiabatic algorithm: its variational parameters are the coefficients of discrete time evolution terms which are tuned so the time evolution of the system approximates the continuous time evolution from the initial to the final Hamiltonian. The number of discrete time evolution terms is called the number of layers ($p$) of the QAOA circuit. It is known that, in the limit as $p$ approaches infinity, the final state of the QAOA circuit approaches the optimal state of the problem \cite{qaoa}. The QAOA circuit takes the form
\begin{align}
\label{eq:qaoa_ansatz}
\ket{\psi(\alpha, \beta)}
=
\left(\prod_p e^{-i\beta_pH_M} e^{-i\alpha_pH_C}\right)
\ket{\psi_{\text{init}}},
\end{align}
where $H_C$ is the cost Hamiltonian (encoding the minimization problem), $H_M$ is a mixing Hamiltonian (encoding an easy-to-solve initial Hamiltonian), and $\ket{\psi_\text{init}}$ is the ground state of the mixing Hamiltonian $H_M$.

Here, $\alpha=\{\alpha_p\}$ and $\beta=\{\beta_p\}$ are sets of variational parameters that are tuned via a classical minimization algorithm running on a classical computer while the quantum computer prepares the trial state $\ket{\psi(\alpha, \beta)}$ (Eq. \ref{eq:qaoa_ansatz}). This state is measured in a manner that allows the classical computer to extract the expectation value of the objective Hamiltonian $\bra{\psi(\alpha, \beta)}H_C\ket{\psi(\alpha, \beta)}$. Minimizing this expectation value minimizes the cost function of the minimization problem.

\section{Homogeneous Sphere Packing}

Sphere packing is an inherently continuous problem in that the spheres may be placed anywhere in space. One way to approximate this continuity (as no computer has infinite memory) is to encode the position of each sphere $s=(x, y, z)$ into the bits or qubits of the hardware. This allows for a great deal of flexibility over the system. The downside is that the number of bits/qubits required to encode the properties scales exponentially with the desired precision. The formulation approaches the continuous case as the precision goes to infinity. The other way to approximate continuity is to discretize space into a grid of points, each of which is assigned a bit/qubit whose value (1 or 0) indicates the existence (or non-existence) of a sphere placed at that location, respectively. This is a less flexible formulation but one whose bit/qubit number scales only polynomially with desired precision. Because of its amenability to QAOA, this second formulation is chosen.

\subsection{Formulation}
\label{src_formulation}

To simplify the problem to a level that can be run on today's hardware, instead of packing spheres into a cylinder, consider packing circles into a circle representing a two-dimensional cross-section of said cylinder. We can then formulate packing as maximizing the number of circles (each of radius $r$) that can be placed within the boundary circle of radius $R_b$ (the radius of the cylinder) such that none of the circles overlap. Note that since all the circles have the same radii, maximizing the number of circles packed is equivalent to maximizing packing density. This problem can be mapped to the problem of Maximum Independent Set (MIS) via the following construction: place an evenly spaced grid of nodes $V$ within the boundary circle and assign edges $v, w \in E$ between all pairs of nodes $v, w \in V^2$ such that circles centered at them would overlap. The integer program for this formulations is
\begin{align}
\label{eq:mis}
\min &\sum_{v\in V} \bar{x}_v \\
\nonumber
\text{s.t.} &\sum_{v,w \in E} x_vx_w = 0,
\end{align}
where $\bar{x}_v = 1 - x_v$ and each indicator variable $x_v$ is 1 or 0 if there is or isn't a circle centered on node $v$, respectively. The objective is the sum of the inverses of the indicator variables over all of the nodes. This represents the number of nodes without circles centered on them, which should be minimized. The constraint encodes that the sum of the product of the indicator variables between every pair of nodes which share an edge must be zero. Note that this can only be satisfied if, for every pair of nodes which share an edge, both are zero. This implies that no two spheres may exist if they would overlap.

\begin{figure}[t]
    \centering
    \includegraphics[width=1\linewidth]{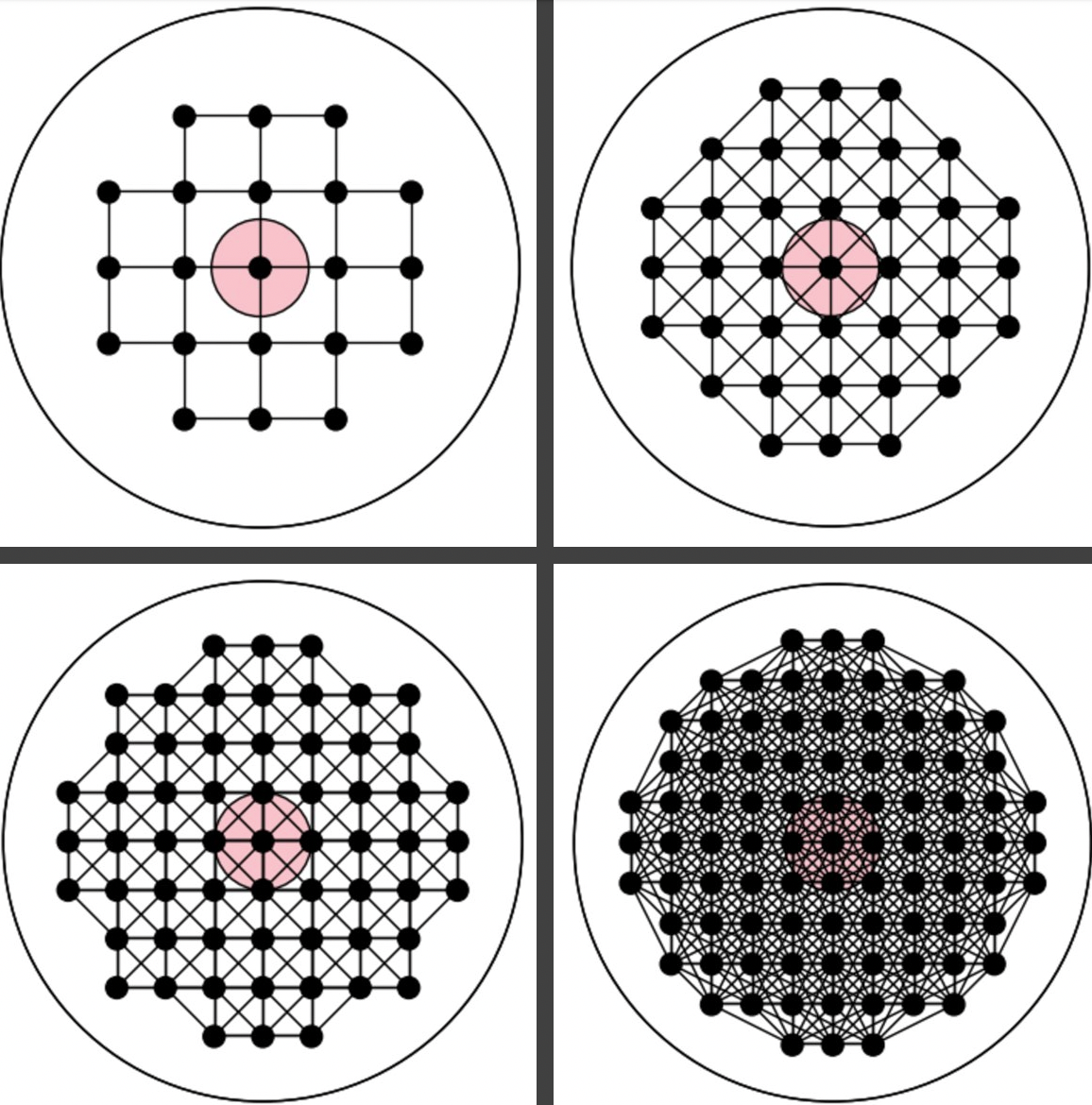}
    \caption{Graphs for discretization spacings $a=$ 1.4, 1.1, 0.9 and 0.75. A red circle of radius $r=1$ is shown in the center of each boundary circle for reference.}
    \label{fig:graphs}
\end{figure}

Figure \ref{fig:graphs} shows how, for fixed radius $r$, the connectivity of the graph increases with decreasing discretization spacing $a$, the spacing between the nodes. As $a$ decreases, the nodes are placed closer together, allowing more of them to fall within $2r$ of each other, at which point they would overlap, triggering the existence of an edge. 

Note that this two dimensional formulation may be extended to the full three dimensional formulation (packing spheres into a cylinder) using the same integer program (Eq. \ref{eq:mis}) by stacking these two-dimensional slices on top of one another and adding additional edges between pairs of points from different slices that are closer than $2r$ to each other. One might be concerned that the connectivity of the graphs for the three dimensional case grows faster than polynomial with the number of points. However, as shown in Section \ref{resource_scaling}, this growth can be bounded by a low degree polynomial.

\subsection{Classical Solution}

\begin{figure}[t]
    \centering
    \includegraphics[width=1\linewidth]{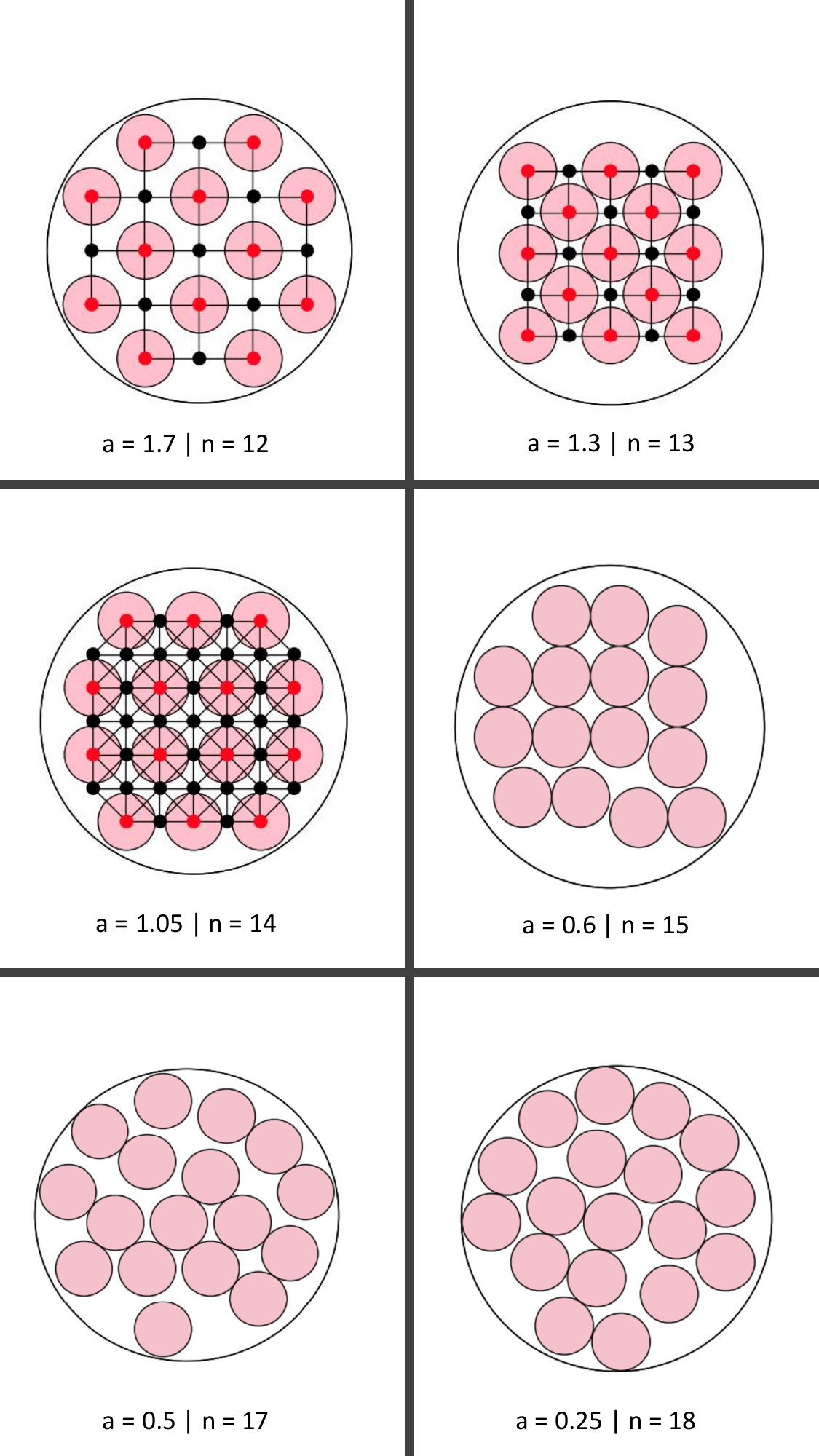}
    \caption{Schematic showing how the size $n$ of the maximum independent set of the graphs increases with decreasing spacing $a$. Note that the underlying graphs have been omitted for the latter half of the examples for clarity of visualization.}
    \label{fig:spacing}
\end{figure}

\begin{figure}[t]
    \centering
    \includegraphics[width=1\linewidth]{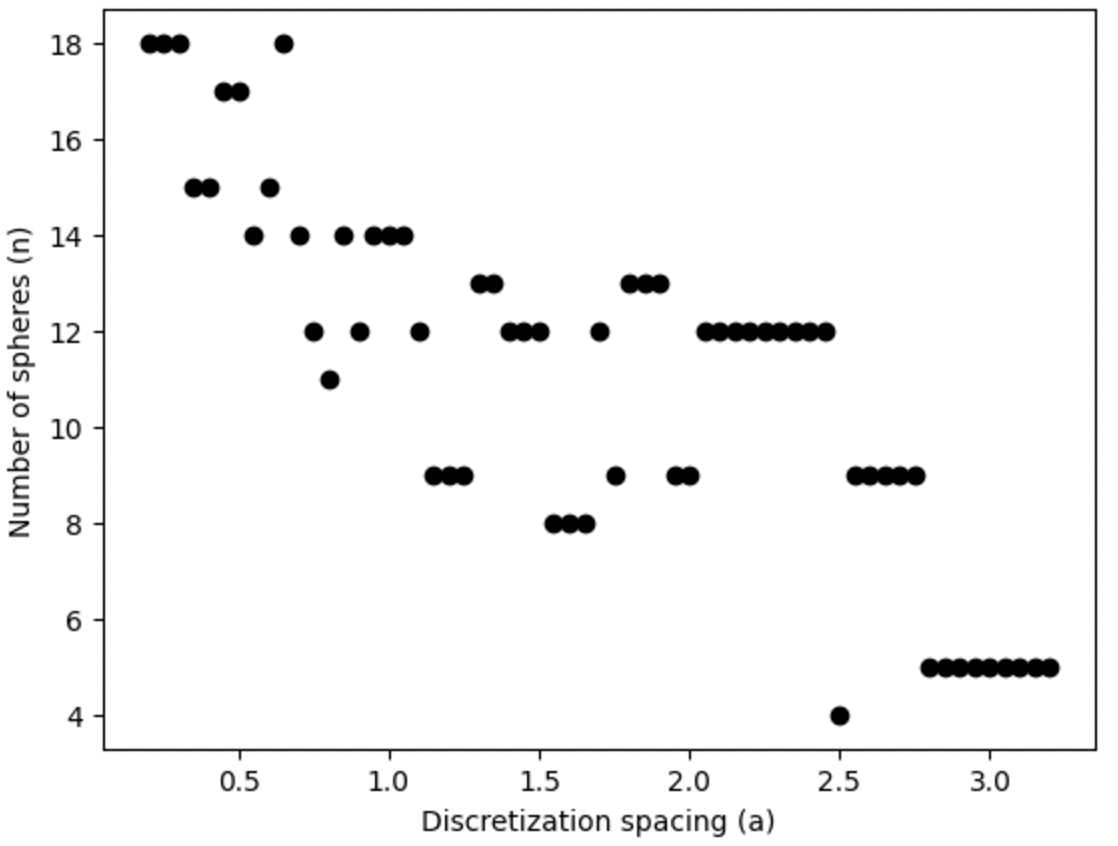}
    \caption{Graph showing that the number of circles packed generally increases with decreasing spacing.}
    \label{fig:n_vs_a}
\end{figure}

As we have formulated our problem as MIS (Eq. \ref{eq:mis}), it can now be approximately solved via classical computation. 
Figure \ref{fig:spacing} shows the maximum independent set found via classical computation for various spacings. It indicates that as the discretization spacing $a$ decreases, the solver is able to pack larger numbers of circles $n$ within the boundary circle (corresponding to higher packing densities). This is because, in the limit as $a$ goes to 0, the formulation approaches the continuous version of circle packing.

Figure \ref{fig:n_vs_a} shows how the number of packed circles scales with the discretization spacing $a$. It appears linear for this discretization spacing interval, though we know that it ultimately must asymptotically approach the maximum number of packable spheres. Therefore, we see that classically, because MIS is NP-hard, it takes an exponential increase in time to achieve a diminishing increase in the number of circles packed. However, for a quantum computer, the resources required to simulate this formulation (Subection $\ref{resource_scaling}$) only grow polynomially.

\subsection{Quantum Solution}

\subsubsection{Hamiltonian Formulation}

To formulate MIS (Eq. \ref{eq:mis}) as a Hamiltonian, we make the following transformation from the indicator variable to Pauli spin matrices
\begin{align}
\label{trans}
x_v \to (I_v - Z_v)/2,
\end{align}
where $I=(\ket{0}\bra{0}+\ket{1}\bra{1})/2$ and $Z=(\ket{0}\bra{0}-\ket{1}\bra{1})/2$, implying $x_v = \ket{1}\bra{1}$. Therefore, $x_v\ket{1}_v = 1$ while $x_v\ket{0}_v = 0$, as expected from an indicator variable. Applying this transformation (Eq. \ref{trans}) to the integer program for MIS (Eq. \ref{eq:mis}) results in the following cost Hamiltonian
\begin{align}
\label{eq:h_hardware}
H_C
=
\frac{1}{2}\sum_{v\in V}Z_v
+
\frac{\lambda}{4}\sum_{v,w\in E}(Z_vZ_w - Z_v - Z_w),
\end{align}
after dropping constant terms, as they only introduce global phase during time evolution. Notice that we have incorporated the constraints via the Lagrange multiplier method by multiplying the constraint Hamiltonian by a hyperparameter $\lambda$ and placing it in summation with the objective Hamiltonian. Additionally, we have selected the $X$ mixer
\begin{align}
H_M = -\sum_{n}X_n,
\end{align}
whose minimum eigenstate is simply $\ket{+}^{\otimes n}$
which can be prepared via the initial application of Hadamard gates $H$ on all qubits in the all zero state.

\subsubsection{Experiment Setup}

\begin{figure}[t]
    \centering
    \includegraphics[width=0.65\linewidth]{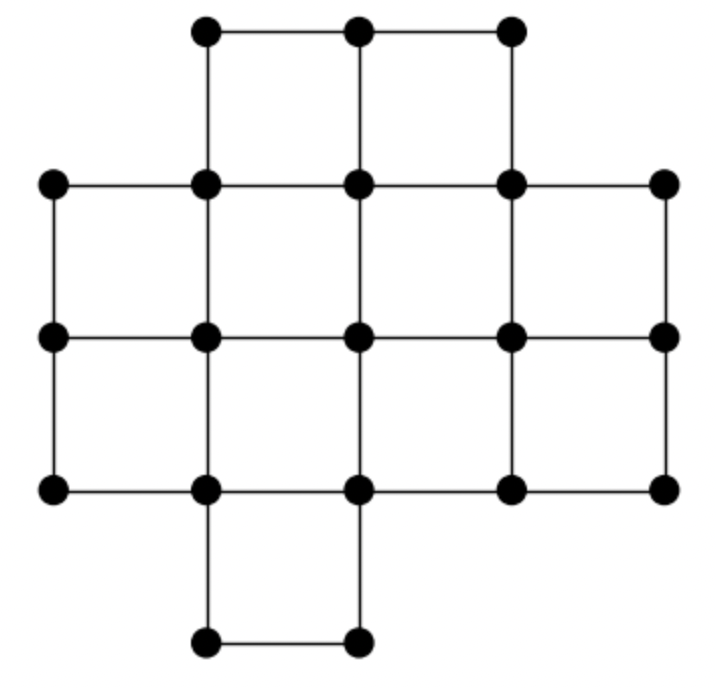}
    \caption{Grid qubit coupling map of IQM's Garnet.}
    \label{fig:garnet_coupling}
\end{figure}

\begin{figure}[t]
    \centering
    \includegraphics[width=0.8\linewidth]{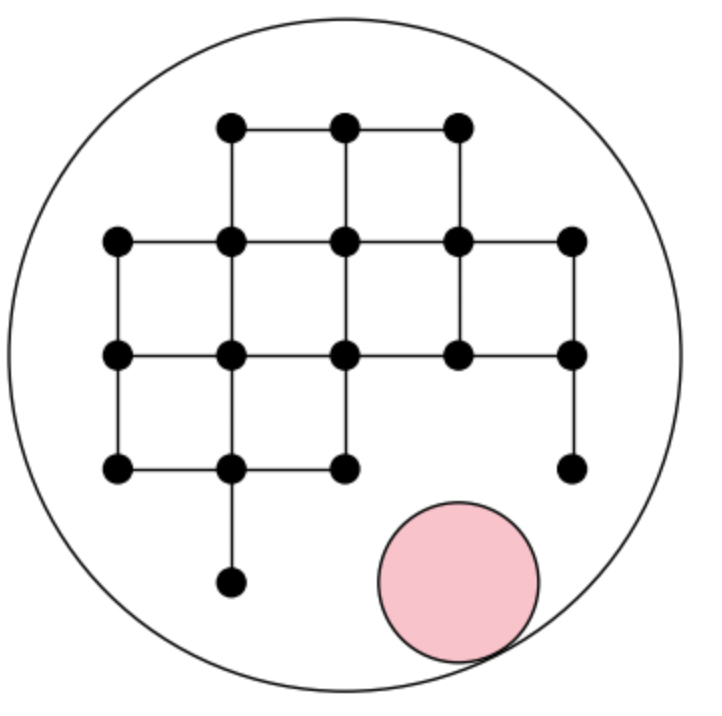}
    \caption{Graph for quantum hardware experiment with first circle touching the  boundary.}
    \label{fig:pictograph}
\end{figure}

\begin{figure}[t]
    \centering
    \includegraphics[width=0.8\linewidth]{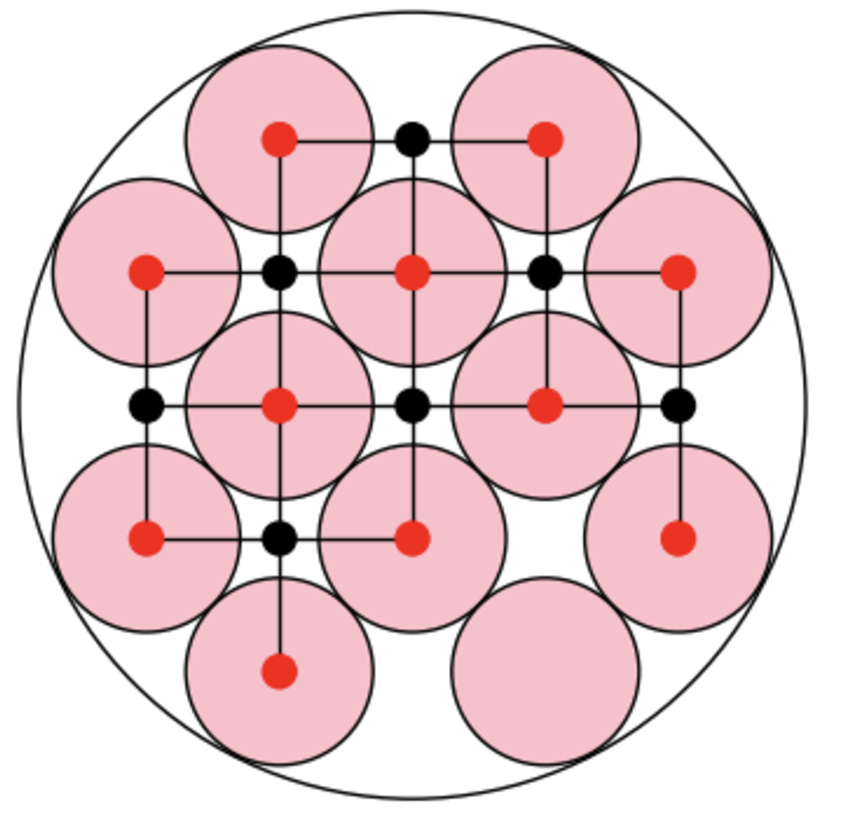}
    \caption{Maximum independent set of the graph for the quantum hardware experiment (Figure \ref{fig:pictograph}). Selected nodes are colored red while non-selected nodes are colored black. Circles are placed and centered at all selected nodes.}
    \label{fig:pictograph_mis}
\end{figure}

The hardware selected for this experiment was IQM's 20-qubit superconducting device, Garnet \cite{garnet}. It was chosen, in large part, because of the grid connectivity of its qubit coupling map (Figure \ref{fig:garnet_coupling}). It's ``missing corners" which prevent it from being a full 5 by 5 grid actually make it a natural fit for circle packing inside a larger boundary circle, as circles placed at said ``missing corners" would extend beyond the boundary. Additionally, the ``missing qubit" at the bottom right-hand corner that would have made the coupling map symmetric is also unnecessary for the problem as any packing of circles may be universally shifted any one direction until at least one of the circles touches the boundary. Therefore, without loss of generality, we start our packing by placing one circle at the location of the ``missing qubit" along the boundary (Figure \ref{fig:pictograph}). Furthermore, the existence of this starting circle precludes the placement of circles at the node directly above it and the node directly to its left. This can be seen when comparing the graph of Garnet's qubit coupling map (Figure \ref{fig:garnet_coupling}) and the graph for the experiment (Figure \ref{fig:pictograph}).
This allows us to reduce the 21 qubits of the naive graph (Figure \ref{fig:garnet_coupling} plus the one ``missing" node) to the 18 nodes shown in Figure \ref{fig:pictograph}. Here, circles have radius $r=1$ and the boundary circle has radius $R_b=4.2$. The graph corresponds to a discretization spacing of $a=\sqrt{2}$. Figure \ref{fig:pictograph_mis} shows that the optimal solution (the maximum independent set) for this discretization spacing is a placement of 12 spheres. This leads to a packing density (fraction of the boundary circle's area filled by packing circles) of $~0.68$. It has been proven that the optimal packing density for 12 spheres $~0.74$ (which has a boundary circle of radius $~4.02$ \cite{best_12}).

\subsubsection{Hyperparameter Optimization}

\begin{figure}[b]
    \centering
    \includegraphics[width=1.0\linewidth]{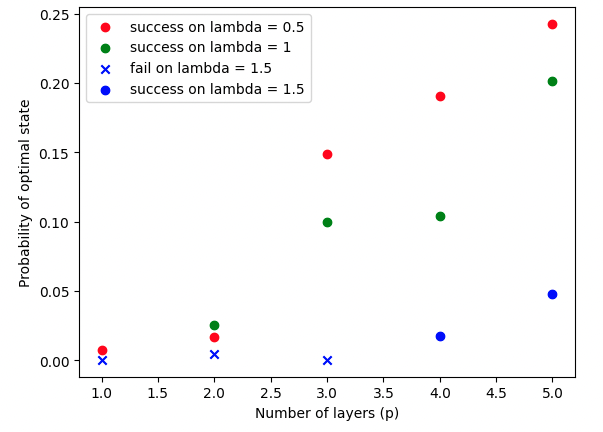}
    \caption{The probability of measuring the optimal state versus the number of layers for various $\lambda$. A circle ($\bullet$) or cross ($\times$) indicates that the optimal state was or wasn't the most probable state measured, respectively.}
    \label{fig:hyper}
\end{figure}

The Lagrange multiplier $\lambda$ in the integer program for MIS (Eq. \ref{eq:h_hardware}) must be chosen carefully. If $\lambda$ is too small, the energies associated with invalid packing will not be sufficiently separated from valid packings. This will lead to QAOA including too many infeasible states in its search, thus hampering both its search efficiency and accuracy. However, if $\lambda$ is too large, it will result in wild swings in energy for the classical minimization algorithm, hampering its ability to minimize the cost landscape. To find a $\lambda$ that achieves the ``sweet spot" between these two extremes, we treat $\lambda$ as a hyperparameter, which is then optimized via a hyper-loop. To objectively judge how well a given $\lambda$ helps QAOA, we plot the probability that QAOA will return the optimal packing vs number of layers $p$ for various $\lambda$ (Figure \ref{fig:hyper}). The plot shows that $\lambda=0.5$ yields the highest probability of finding the optimal state as $p$ grows while additionally finding the optimal state with the highest probability for all $p$. Thus $\lambda=0.5$ is chosen for running QAOA on real quantum hardware.

Given the increasing optimal state overlap with decreasing $\lambda$, one may wonder why we don't select and even smaller $\lambda$. However, further numerical simulations indicated that for smaller $\lambda$'s, namely $\lambda=0.25$, the energy penalty for invalid packings was so small that, even after 20,000 runs of a noiseless simulation, QAOA couldn't find the optimal solution even once (across all $p$). This would constitute an example of $\lambda$ being too small, as the small energy penalty allowed invalid packings to swamp the optimizer, precluding it from finding the optimal solution.

\subsubsection{Compilation}
\label{subsec:compilation}

\begin{figure}[t]
    \centering
    \includegraphics[width=0.65\linewidth]{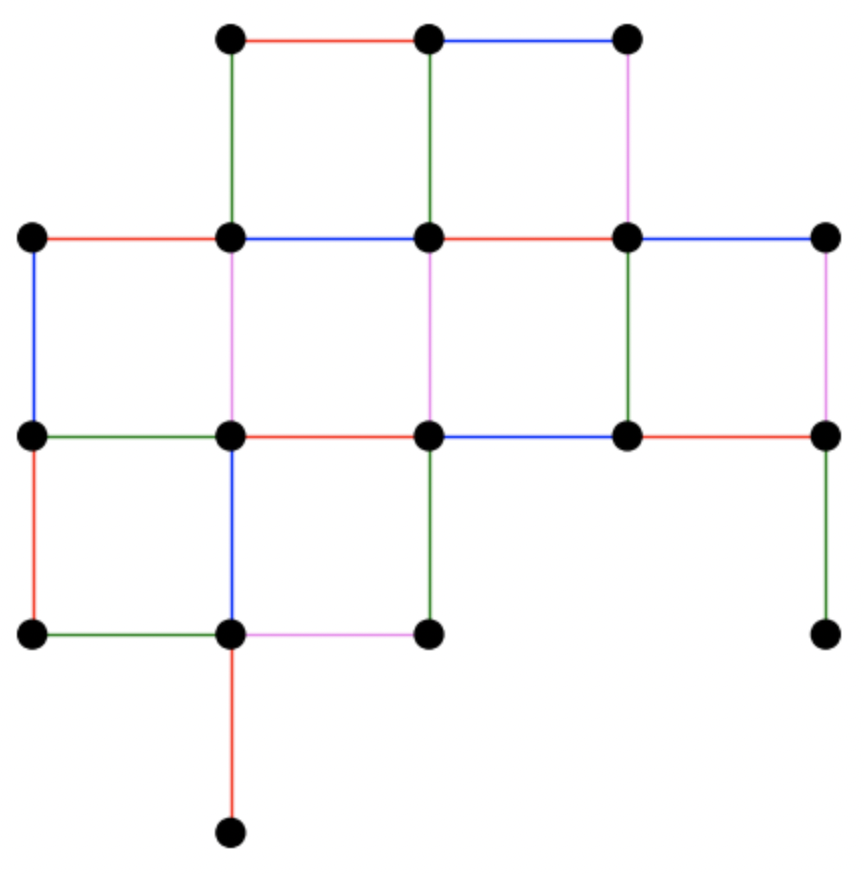}
    \caption{Maximum edge coloring of the Garnet coupling map. The $\text{R}_{ZZ}$ gates between edges of the same color are applied simultaneously.}
    \label{fig:coloring}
\end{figure}

Plugging the MIS Hamiltonian (Eq. \ref{eq:h_hardware}) into the QAOA ansatz (Eq. \ref{eq:qaoa_ansatz}) results in a set of $\text{R}_{ZZ}$, $\text{R}_Z$, and $\text{R}_X$ gates. The $\text{R}_{ZZ}$ operation can be decomposed into single-qubit gates and exactly two CZ gates (Garnet's native two-qubit gate). Meanwhile, to reduce total gate count, the $\text{R}_Z$ and $\text{R}_X$ gates can be combined (as $\text{R}_Z$ commutes through $\text{R}_{ZZ}$) and further expressed in terms of $\text{PR}_X$ gates (phased-$\text{R}_X$ gates, IQM's native single qubit gate).
Furthermore, in an effort to minimize the depth of the quantum circuit, we partition the $\text{R}_{ZZ}$ gates into 4 disjoint sets, corresponding to the maximum edge coloring of the coupling map (Figure \ref{fig:coloring}). Because no edges of the same color share a node, every $\text{R}_{ZZ}$ corresponding to an edge of the same color may be applied simultaneously. Thus, all $\text{R}_{ZZ}$ gates may be applied with only depth 4, the smallest possible number of disjoint sets for this graph.

\subsubsection{Noisy Simulation}
\label{subsec:noisy_sim}

\begin{figure}[t]
    \centering
    \includegraphics[width=1.0\linewidth]{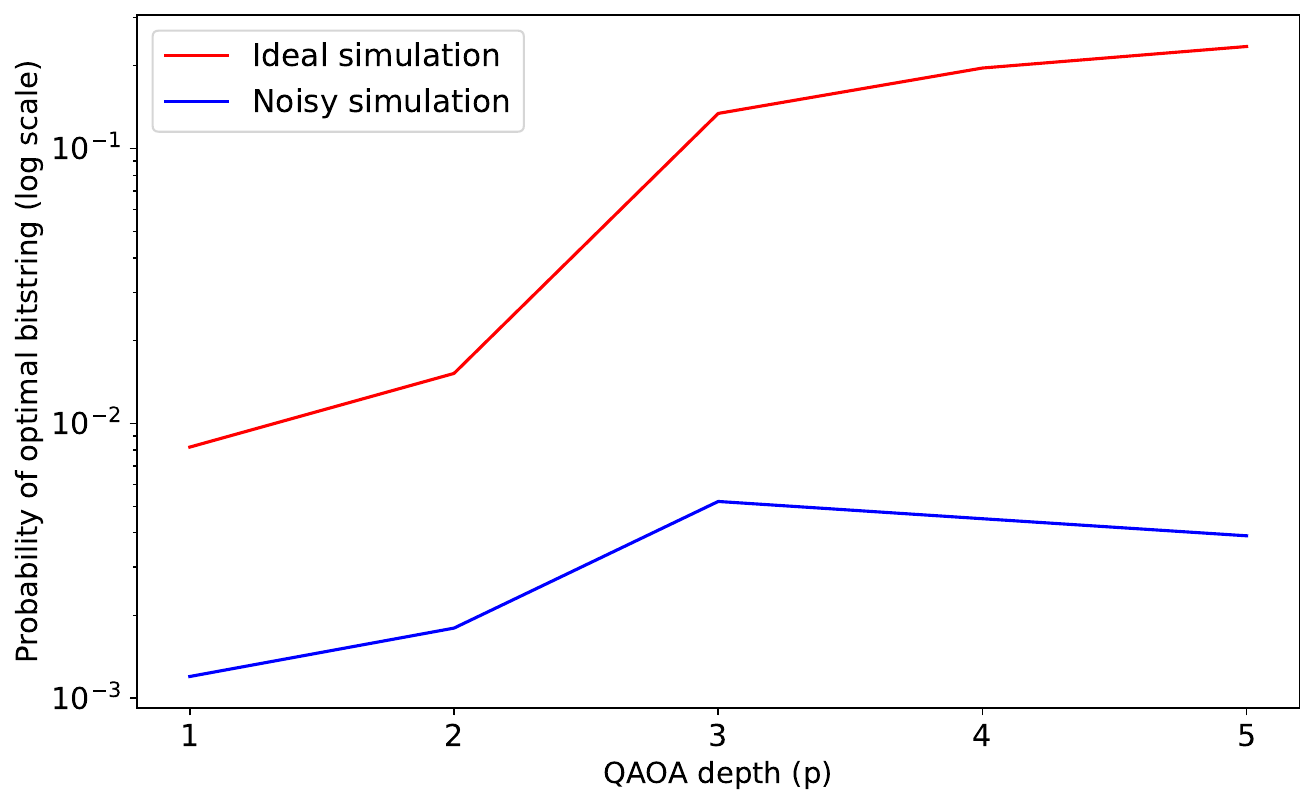}
    \caption{Plot of an ideal and noisy simulation of compiled QAOA circuits with increasing depth $p$. The simulations were done by sampling 20,000 shots, using pre-computed angles, and setting $\lambda=0.5$. The noisy simulation was subject to single-qubit, two-qubit, and readout errors which model IQM's Garnet.}
    \label{fig:ideal_noisy_sim_result}
\end{figure}

Before executing our QAOA circuits on hardware, we considered an ideal simulation of the QAOA circuit with $\lambda=0.5$ over increasing circuit depth $p$ and found that, as expected, the probability of QAOA measuring the optimal solution increases with increasing depth (Figure \ref{fig:ideal_noisy_sim_result}). However, because we are currently in the Noisy Intermediate-Scale Quantum (NISQ) \cite{preskill2018quantum} era, the existence of a consistent level of uncorrected noise makes obtaining good results difficult as circuit depth increases. That is, there is a balance to be had (at least for NISQ devices) in the circuit depth that can be run before the increasing noise from gate count overwhelms QAOA from finding the optimal solution with the highest probability. To this end, we additionally ran a noisy simulation of our QAOA circuits to predict which depth $p$ would perform best on the selected hardware (tailoring our noise parameters to those of IQM’s Garnet).

We started by optimally compiling the circuit (Subsection \ref{subsec:compilation}) thus reducing the need to consider noise arising from circuit routing and its resultant introduction of noisy SWAP gates. We then modeled single- and two-qubit errors in addition to measurement readout errors. The device noise calibration data were obtained from AWS Braket shortly before the time of the experiment. The single-qubit errors consisted of an amplitude damping channel based on the $T_1$ thermal relaxation times reported for each single qubit, and a phase damping channel on each qubit following its respective $T_2$ de-phasing time. Lastly, we included a single-qubit depolarizing channel with the depolarization rate taken to be roughly as one minus the fidelity of each respective qubit’s simultaneous randomized benchmarking result. Given that the only two-qubit gate in our QAOA circuits is the CZ gate, we employed a two-qubit depolarizing noise channel to model that gate error source – with the depolarization rate taken to be the reported CZ infidelity per qubit pair in the coupling map. Finally, we modeled the expected readout error of measurements on hardware via a bit flip channel with a probability of one minus the respective qubit's readout rate. 
Executing a noisy-simulation using all of these channels per gate operation resulted in the blue line of in Figure \ref{fig:ideal_noisy_sim_result}. Following our earlier expectation, we find that the various noise sources lower our probabilities and we also understandably find that, after $p=3$, the probability of the optimal solution actually decreases – contrary to the ideal case scenario. 

This highlights that the limitation of running on quantum hardware stems more from their current noise levels than from the ability of QAOA to perform effectively. We therefore expect that continued reduction in hardware noise, along with a combination of quantum error mitigation, suppression, and correction, will help overcome this limitation. In the current term though, we see that carrying out a noisy simulation can help inform the circuit depths that could be run on hardware before qubit noise becomes a significantly diminishing factor, as we have done here.

\subsubsection{Experiment Results}

To execute our circuits on Garnet, we compiled and submitted our jobs through Infleqtion's compilation software, Superstaq \cite{superstaq} which, in turn, submitted the job through AWS's Braket. We ran QAOA for $p=3$ (based on the noisy simulations of Subsection \ref{subsec:noisy_sim}) and also $p=1$ and $p=5$ for comparison. Each QAOA circuit was executed with 20,000 shots (the maximum number of shots allowed by IQM for submissions to Garnet through AWS). The variational parameters $\alpha$ and $\beta$ were trained via classical simulation and the final quantum circuit was run on the quantum device using these optimized parameters.
In the small case of this example experiment, classical computers can simulate the quantum circuit; however, at larger problem sizes, one may take advantage of parameter concentration (Section \ref{param_conc}) to train the variational parameters via classical simulation on a smaller, classically trainable, sub-problem and then run the quantum circuit for the full problem with these optimized parameters on a quantum computer. At a large enough scale, the quantum computer becomes necessary for running the final circuit as classical simulation of quantum circuits scales exponentially with the number of qubits.

\begin{figure}[t]
    \centering
    \includegraphics[width=1\linewidth]{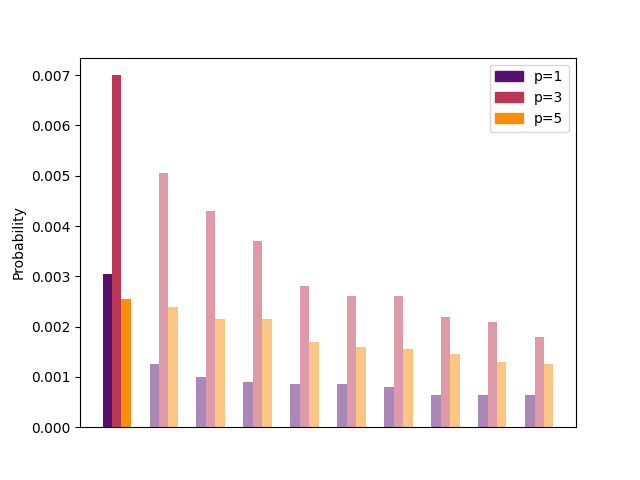}
    \caption{Probabilities of the top 10 most probable states (rank ordered) measured on Garnet for $p=$ 1, 3, and 5. The opaque bars represent the optimal solution while the transparent bars represent non-optimal solutions.}
    \label{fig:plot_hist}
\end{figure}

Figure \ref{fig:plot_hist} shows the probabilities of the top ten most probable states (rank ordered) measured on Garnet for various $p$. The opaque bars represent the optimal solution while the translucent bars represent non-optimal solutions. Note that for each $p$, the most probable state measured was also the optimal state. The figure also demonstrates the two competing forces involved when selecting the best $p$. On the one hand, increasing $p$ increases the expressiveness of the circuit and therefore the effectiveness of QAOA. On the other hand, increasing $p$ also increases the depth of the quantum circuit, and therefore the noisiness of the results. For example, the plot for $p=1$ is sharp because of its short depth, yet low-probability because of its low expressivity. By contrast, the plot for $p=3$ is high probability because of its high expressivity, yet flat because of its long depth. The `sweet spot' is at $p=3$ (as suggested by our noisy simulations from Subsection \ref{subsec:noisy_sim}), where the circuit is expressive enough to render high probabilities yet short enough to be sufficiently sharp. As one expands to larger, more expensive-to-run problem sizes, the importance of preemptively searching for the ``sweet spot" for $p$ via noisy simulations of the circuit before running on hardware (Subsection \ref{subsec:noisy_sim}) becomes ever more apparent.

\subsubsection{Parameter Concentration}
\label{param_conc}

\begin{figure}[t]
    \centering
    \includegraphics[width=.8\linewidth]{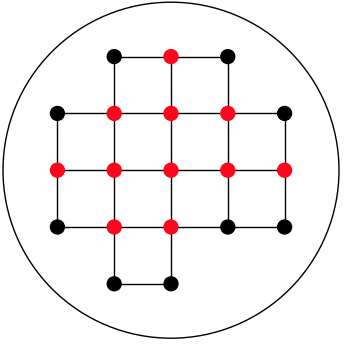}
    \caption{Full graph (all dots) and subgraph (red dots) used for the parameter concentration study.}
    \label{fig:sub_graph}
\end{figure}

\begin{figure}[b]
    \centering
    \includegraphics[width=1\linewidth]{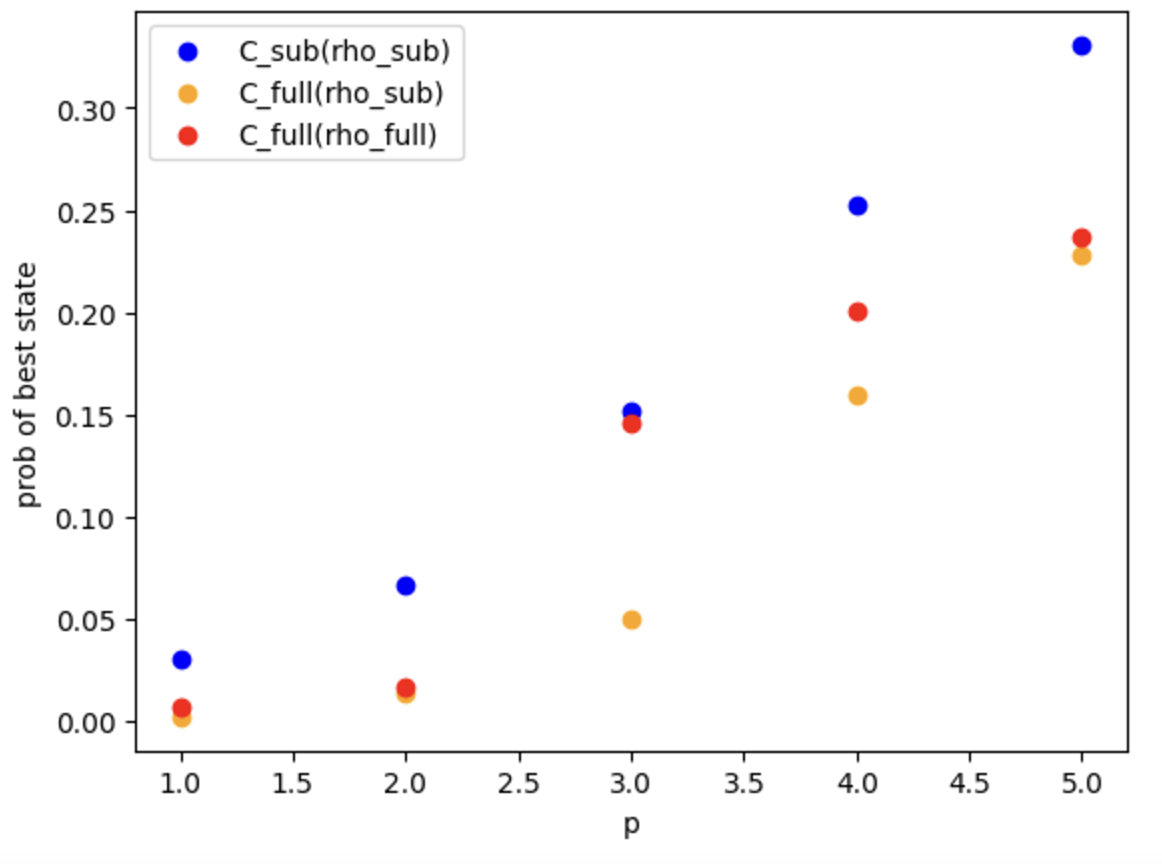}
    \caption{Probability of measuring the best state versus number of layers $p$ for three instances: 
    the subgraph solved on subgraph trained parameters $C_{\text{sub}}(\rho_{\text{sub}})$, 
    the full graph solved on subgraph trained parameters $C_{\text{full}}(\rho_{\text{sub}})$, 
    the full graph solved on full graph trained parameters $C_{\text{full}}(\rho_{\text{full}})$.}
    \label{fig:param_conc_prob}
\end{figure}

% \begin{figure}[h]
%     \centering
%     \includegraphics[width=1\linewidth]{param_conc_ar.png}
%     \caption{}
%     \label{fig:param_conc_ar}
% \end{figure}

As noise often scales with the number of qubits/gates, and therefore problem size, it becomes harder to train QAOA as the problem size increases. 
This roadblock can be alleviated, however, if the problem exhibits parameter concentration \cite{param_conc}, a phenomenon in which the QAOA parameters trained on a smaller instance of a problem perform well when applied to a larger instance of the same problem.
Fortunately, we have numerical evidence suggesting parameter concentration for our particular problem of sphere packing.
Figure \ref{fig:sub_graph} shows the full graph used for the hardware experiment. The dots colored red represent a subgraph which will serve as the ``smaller problem instance" for our parameter concentration investigation. 
We train the parameters of both the subgraph QAOA circuit ($C_{\text{sub}}$) and the full graph QAOA circuit ($C_{\text{sub}}$), resulting in $\rho_{\text{sub}}$ and $\rho_{\text{full}}$, respectively. 
To test for parameter concentration, we compare the performance of the full graph circuit executed with the parameters trained on the subgraph $C_{\text{full}}(\rho_{\text{sub}})$ with that of the full graph circuit executed with the parameters trained on the full graph $C_{\text{full}}(\rho_{\text{full}})$. Figure \ref{fig:param_conc_prob} compares the performance of $C_{\text{full}}(\rho_{\text{sub}})$ and $C_{\text{full}}(\rho_{\text{full}})$ with respect to overlap with the optimal state for various numbers of layers $p$. It also includes the performance of the subgraph circuit executed with the parameters trained on the subgraph $C_{\text{sub}}(\rho_{\text{sub}})$. We can see that QAOA yields a larger overlap with the optimal state when solving the subgraph with the parameters trained on the subgraph $C_{\text{sub}}(\rho_{\text{sub}})$ when compared to solving the full graph with the parameters trained on the full graph $C_{\text{sub}}(\rho_{\text{sub}})$. This is expected as the subgraph has less parameters and is therefore easier to train. However, we can also see that QAOA on the full graph with parameters trained on the subgraph $C_{\text{full}}(\rho_{\text{sub}})$ performs only slightly worse than QAOA on the full graph with parameters trained on the full graph $C_{\text{full}}(\rho_{\text{full}})$, demonstrating parameter concentration. This suggests that QAOA may be able to solve real-world instances of this problem that are too large to train by executing QAOA on parameters trained on a trainable sub-instance.

\section{Heterogeneous Packing}

Using the encouraging results from real quantum hardware as a springboard, we generalize sphere packing to the heterogeneous case, for which good classical solutions are much harder to achieve. Consider the case of $s+1$ distinct radii $R = \{r_n\}_{n\in N}$ where $N=\{0, 1, ..., s\}$ and $r_0=0$ (a useful indicator for the nonexistence of a sphere). The state of the qubit (or set of qubits) at each node informs which sized sphere is centered at said node. To maximize the density of the packing, we minimize empty space subject to the constraint that the arrangement of spheres is non-overlapping.
We formulate the cost Hamiltonian $H_C = H_{obj} + \lambda H_{con}$ for heterogeneous packing via two different formulations, which we call ``First Quantization" and ``Second Quantization" because of their analogies to the two levels of quantization in physics.

% $H_{obj}$ is the objective Hamiltonian, the minimization of which encourages the packing to maximize the volume taken up by the spheres. 
% $H_{\text{con}}$ is the constraint Hamiltonian, which adds an energy penalty for every pair of overlapping spheres within the packing.
% $\lambda$ is a positive real number (akin to a Lagrange multiplier) chosen to ensure that non-valid packings have higher energies than valid ones. In practice, $\lambda$ should be small enough to keep the energy scales of $H_{obj}$ and $\lambda H_{con}$ close enough to prevent massive jumps in $\langle H \rangle$ during training, yet large enough to penalize the energies of non-valid packings enough such that the optimizer can find sufficiently optimal, valid states.
% A (non-strict) lower bound on $\lambda$ ensuring all the energies of non-valid packings are greater than all the energies of valid packings is $\sum_{v\in V}\sum_{r\in R_v}V(r)$, since this is the maximum reward of the objective Hamiltonian. $V(r)$ is simply the volume of the sphere with radius $r$.

\subsection{``First Quantization" Formulation}

\begin{figure}[t]
    \centering
    \includegraphics[width=1\linewidth]{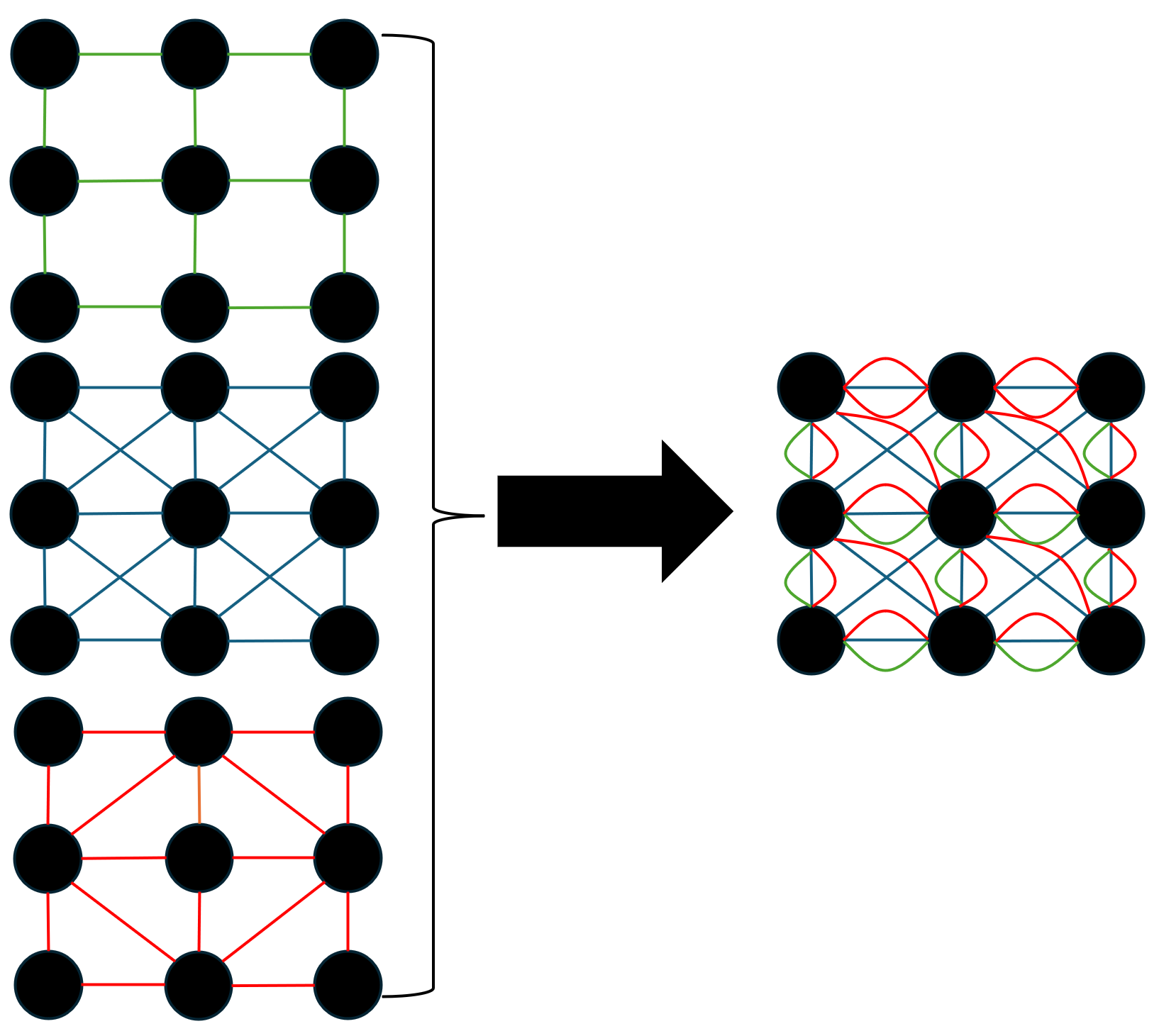}
    \caption{For the ``First Quantization" formulation of the problem, we assume that there is one set of nodes $V$ with several edge sets for the restrictions on the placement of pairs of different sized spheres, which can be represented as different edge colors. Each node is then colored as well based on which sized sphere is placed there. This structure allows for a logarithmic compression in the number of qubits in exchange for a Hamiltonian with higher-order terms.}
    \label{fig:color2}
\end{figure}

In the ``First Quantization" formulation, we assign a qudit state $\ket{n}_v$ to each vertex $v\in V$ to indicate that the radius of the sphere centered at vertex $v$ is $r_n$. Recall that we always set $r_0=0$ to represent that no sphere is placed at the node because this is equivalent to placing a sphere of radius zero at the node. 
This formulation is ``first-quantized'' in the sense that a property (the radii) of a fixed grid of spheres are encoded into the states of the qudits.
Written below are the ``First Quantization" formulated Hamiltonians 
\begin{align}
H_{\text{obj}} 
&=  
-\sum_{v\in V}\sum_{n \in N_v}
V(r_n)\ket{n}\bra{n}_v
\\
H_{\text{con}}
&= 
\sum_{v,w\in V}\sum_{n,m \in N_{vw}}
\ket{nm}\bra{nm}_{vw} ,
\end{align}
where
\begin{align}
N_v &= \{n \in N : |v| + r_n \leq R_b\} \\
N_{vw} &= \{n,m \in N_v \times N_w : |v - w| \leq r_n + r_m\}.
\end{align}
Here, $N_v$ represents the index set of the set of radii that admit spheres of said radii to be placed at node $v$ without extending beyond the boundary of the cylinder. 
Similarly, $N_{vw}$ represents the index set of pairs of radii that would not admit the placement of spheres of said radii at nodes $v$ and $w$ without their overlapping.
Here, $R_b$ is the radius of the cylinder. 
Converting these Hamiltonians to a qubit formulation (a requirement for any real hardware system today) requires only $\log_2|R|$ overhead via a binary encoding.

\subsection{``Second Quantization" Formulation}

\begin{figure}[t]
    \centering
    \includegraphics[width=1\linewidth]{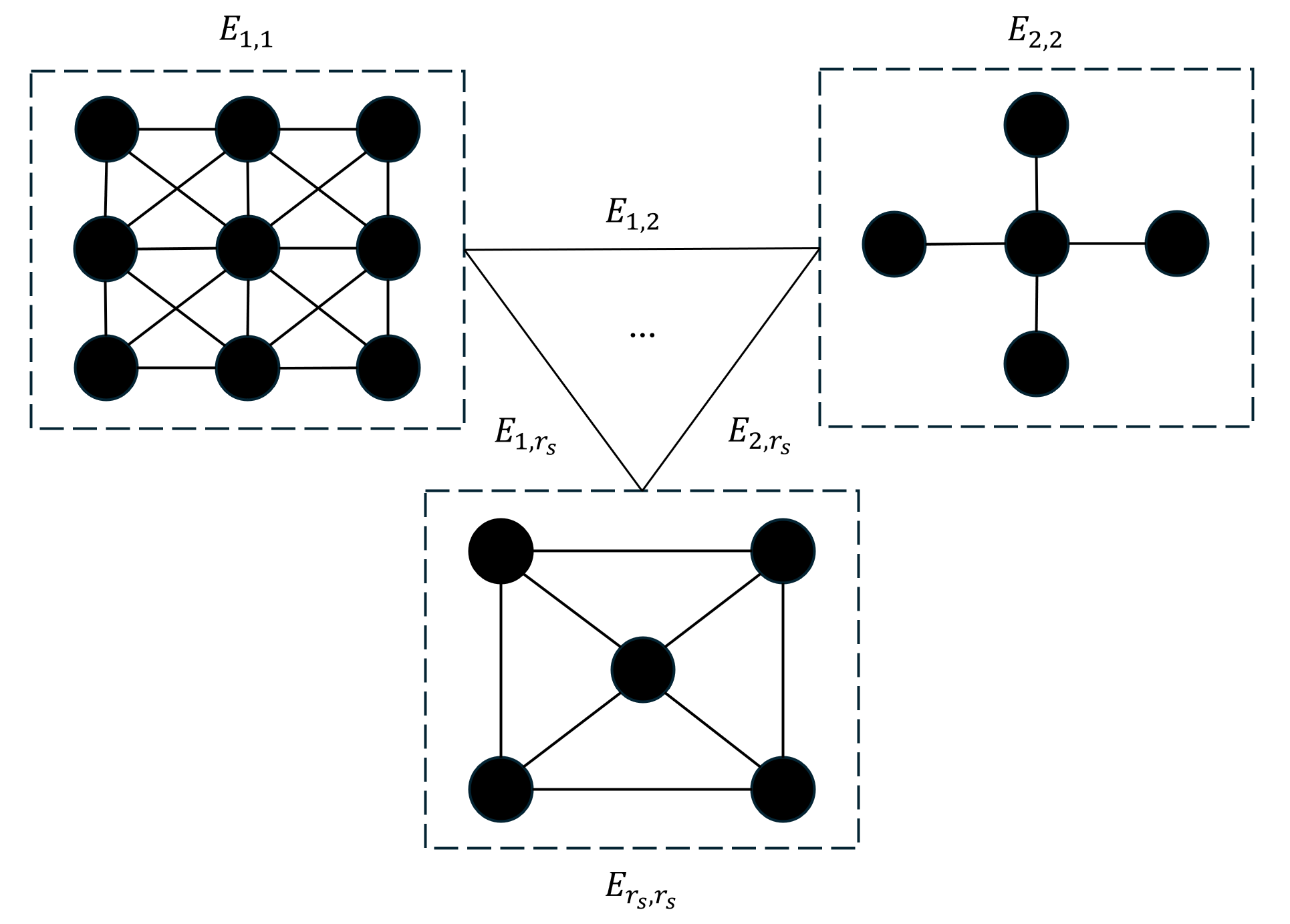}
    \caption{In the ``Second Quantization" formulation of the problem, we loosen the assumption that the set of available placements is the same for all radii. As a result, we build a new graph by stitching together each of the graphs that we would have built in the homogeneous radius case. In this way, the multiple edge sets of the previous case become one edge set over many subgraphs.}
    \label{fig:stitch}
\end{figure}

In the ``Second Quantization" Hamiltonian formulation of the problem, we assign a qubit state $\ket{q}_v$ to each vertex $v\in V_r \subset V$ to indicate the existence ($q=1$) or non-existence ($q=0$) of a sphere of radius $r$ at vertex $v$. This formulation is ``second-quantized" in the sense that properties (the radii and locations) of the spheres are encoded into the indices of the qubits while the spheres' existences are encoded into the states of the qubits. Written below are the ``Second Quantization" formulated Hamiltonians 
\begin{align}
H_{\text{obj}} 
&=
-\sum_{r \in R}\sum_{v \in V_r}V(r)\ket{1}\bra{1}_v
\\
H_{\text{con}} 
&= 
\sum_{rs \in R^2}\sum_{vw\in E_{rs}}\ket{11}\bra{11}_{vw} ,
\end{align}
where 
\begin{align}
E_{rs} = \{v,w \in V_r\times V_s : |v - w| \leq r + s\}.
\end{align}

Figure \ref{fig:stitch} shows qbsolve's \cite{qubosolve} attempt at solving the ``second quantization" formulation of the heterogeneous circle packing problem.

\begin{figure}[t]
    \centering
    \includegraphics[width=1\linewidth]{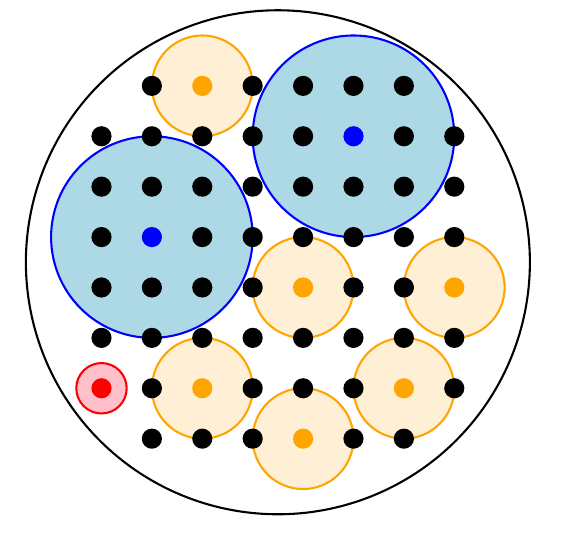}
    \caption{Numerical simulation using qbsolve of the ``second quantization" formulation of the heterogeneous circle packing problem with three distinct allowable radii.}
    \label{fig:stitch}
\end{figure}

% \subsection{Analytical Results}

% \begin{figure}[h]
%     \centering
%     \includegraphics[width=0.7\linewidth]{multi_sphere_analytical.png}
%     \caption{Analytical result using $p=1$ RQAOA.}
%     \label{fig:mis}
% \end{figure}

% \section{Constant Sphere Number Case}

% In this formulation of the problem, we fix the total number of spheres of each radii and attempt to minimize the distance between their center of mass and an arbitrary point in space. The formulation can be written as the following integer program
% \begin{align*}
% \max &\sum_{r\in R}\sum_{v_r\in V_r}m(r)rx_{v_r}
% \\
% \text{s.t.} &\sum_{r\leq s\in R}\sum_{v_r,w_s\in E_{rs}}x_{v_r}x_{w_s} = 0, 
% \\
% &\Big\{\sum_{v_r\in V_r}x_{v_r} = k_r\Big\}_{r \in R}
% \end{align*}
% where $m(r)$ and $k_r$ are the mass and number of spheres of radius $r$, respectively.

% \section{Alternative Approaches}

% \subsection{``First Quantized" Analog Quantum Formulation}

% \subsection{``Second Quantized" Analog Quantum Formulation}

\section{Resource Scaling}
\label{resource_scaling}

% To bound the number of qubits and the number two qubit gates in the ``Second Quantized" encoding of the problem, we need to count the number of nodes and edges, respectively. This counting exercise is equivalent to counting the terms in $H_{obj}$ and $H_{con}$. The number of qubits is the number of terms in $H_{obj}$, and two times the number of terms in $H_{con}$ provides an upper bound on the number of two qubit gates needed to implement $e^{\gamma H}$ in the QAOA.
% Using the terminology from Section III, the number of terms in $H_{obj}$ is $\sum_{r\in R}|V_r|$. Supposing each subgraph is a d-dimensional lattice with $q$ points per side, this quantity is upper bounded by $|R|q^d$.
% Similarly, for the number of terms in $H_{con}$ we can consider the worst case scenario for the same d-dimensional lattice. The worst case is when each point in each lattice needs an edge between each other point in the same lattice, as well as each point in each other lattice. The number of such edges would be 

Resource estimation for this problem is important to argue for the potential for quantum advantage over classical computation. To estimate this scaling of quantum resources for each problem formulation, we start by making several reasonable assumptions.
Suppose we have a $d$-dimensional lattice with $q$ points per side placed within a $d$-dimensional cylinder with radius $R_b$. Furthermore, notate the set of radii, node sets, edge sets and the maximum radius as
\begin{align}
    R&=\{r_1,...,r_s\} \\
    \nonumber
    V&=\{V_{r_1},...,V_{r_s}\}\\
    \nonumber
    E&=\{E_{r_1 r_1}, E_{r_1 r_2},...,E_{r_s r_s}\}\\
    \nonumber
    r_{m}&=\max_{r\in R}r,
\end{align}
respectively.
For the ``Second Quantization" problem formulation, the number of qubits is
\begin{align}
    \sum_{r\in R}|V_r| &\leq |R|\max_{r\in R}|V_r|
    \nonumber
    \\
    &\leq|R|q^d.
\end{align}
The number of CNOT gates to implement $e^{iH\gamma}$ in one QAOA layer is
\begin{align}
    \sum_{r_i,r_j\in R^2}|E_{r_i r_j}| &\leq |R|^2q^d\max_{r\in R}\left(\max_{v\in V_r} \deg(v)\right)
    \nonumber 
    \\
    &\leq |R|q^d\left(\frac{r_m q}{R_b}\right)^d 
    \nonumber
    \\
    &= |R|\left(\frac{r_mq^2}{R_b}\right)^d.
\end{align}

For the ``First Quantization" problem formulation, we define a similar host of sets:
\begin{align}
    R &= \{r_1,...,r_s\}
    \nonumber 
    \\
    \Pi&=\{1,...,\lceil \log_2\left(|R|+1\right)\rceil\}
    \nonumber
    \\
    V&=\{v\}
    \nonumber
    \\
    E&=\{E_{r_1 r_1}, E_{r_1 r_2}, ..., E_{r_s r_s}\},
\end{align}
where $E_{r_i r_j}$ has an edge $(v,w)$ if and only the distance between nodes $v$ and $w$ is less than $r_i+r_j$. With these definitions, the number of qubits is simply
\begin{align}
    |\Pi||V| &\leq \left(\lceil\log_2\left(|R|+1\right)\rceil\right)q^{d}
    \nonumber
    \\
    & \approx \log|R|q^{d}.
\end{align}
Moreover, to count the number of CNOTs, we start by defining the single state penalty function
\begin{equation}
g(v, r_i, w, r_j)=
    \begin{cases}
        1 & \text{if} \ vw \in E_{r_ir_j}
        \\
        0 & \text{otherwise}
    \end{cases}
\end{equation}
The single state penalty function $g$ can usually be implemented in fewer gates than the worst case, but it will always be bounded by $2^{2|\Pi|}$. Now we can bound the number of CNOTs as
\begin{align}
    \sum_{r_i,r_j\in R^2}\sum_{v,w\in E_{r_i r_j}} &\leq |R|2^{2|\Pi|}\max_{r_i,r_j\in R^2}\left(\max_{v\in V} \deg_{r_i r_j}(v)\right)
    \nonumber
    \\
    &\leq |R|2^{\left({2\lceil \log_2 \left(|R|+1\right\rceil}\right)}\left(\frac{r_mq}{R_b}\right)^d 
    \nonumber
    \\
    & \approx |R|^3\left(\frac{r_mq}{R_b}\right)^d.
\end{align}
Note that the bound on the single state penalty function is quite loose. Additionally, while the scaling may appear exponential in $d$, recall that for real-world applications $d$ is only $2$ or $3$, keeping the quantum resource scaling for both formulations bounded by a low order polynomial in all its components. Note that compared to the ``first-quantization" formulation, the ``second-quantization" formulation trades an increase in the CNOT scaling degree for an exponential decrease in the number of qubits. The best formulation to use is therefore dependent on which resource will incur a larger noise penalty for near-term quantum devices, their number of qubits or number of gates. Either way, we have shown that the quantum resources for this problem only scale polynomially with increased problem size - in contrast to classical computers, whose complexity grows exponentially as this problem is NP-hard.

\section{Conclusion}

In this work, we model a packed chromatography column using a mathematical formulation of sphere packing. We identified three complexity levels of sphere packing. The first level was solved on real quantum hardware, the second was simulated on a classical computer (with a road map to quantum simulations), and the third was shown to be an extension of the first two. Accompanying the quantum hardware results: hyperparameter optimization was conducted to determine the optimal penalty term $\lambda$ for the problem, optimal compilation was implemented to reduce noise, and noisy simulation was conducted to preemptively suggest a number of layers $p$ with which to run QAOA on quantum hardware. Finally, evidence of parameter concentration was found for the problem, suggesting that large, untrainable instances of the problem may still be solvable by QAOA via execution on parameters trained on smaller, trainable, problem instances.
The fact that classical algorithms scale exponentially for this NP-hard problem opens up the possibility of quantum advantage in the near to long term. Robust resource scaling of the quantum resources required to solve the problem on a quantum computer was conducted to further support this claim. In the future, we would like to attempt to solve the ``heterogeneous circle packing" complexity level on a quantum computer in the near-term and the ``heterogeneous sphere packing" complexity level in the long term. Altogether, this work provided a base and road map to potential quantum advantage for the problem of column chromatography modeled as sphere packing, which could have profound positive implications for the future of the biopharmaceutical industry and the health and wellness of the public.

\section{Acknowledgments}

This work was funded by CSL Behring. The authors used Infleqtion's compilation software, Superstaq, to submit circuits to IQM's quantum computer, Garnet, through AWS's Braket.
From Infleqtion, we would like to thank Pranav Gokhale for guidance on quantum hardware selection, Caitlin Carnahan for serving as program manager for this project, and Peter Noell for managing coordination between Infleqtion and CSL. Additionally, we acknowledge Cameron Booker for theoretical insights including the graph formulation for the hardware experiment. 
Additionally, we acknowledge Bao Bach, Rajas Chari, Kabir Dubey, Rohan Mehta, Shivam Mundhra, and Carlo
Siebenschuh for proposing the formulation of chromatography as sphere packing in their submission (mentored by Infleqtion's Victory Omole) at the quantum hackathon Big Q Hack, upon which this work was inspired. 
Finally, we acknowledge the work of every scientist that has come before us
as ``if (we) have seen further, it is by standing on the shoulders of giants" - Sir Isaac Newton.

\bibliography{main.bib}

\end{document}